\documentclass[aj,twocolumn]{aastex61}
\pdfoutput=1%for arXiv submission
\usepackage{amsmath,amstext}
\usepackage[T1]{fontenc}
\usepackage[utf8]{inputenc}
\usepackage{apjfonts} 
\usepackage[figure,figure*]{hypcap}
\usepackage{amsmath,amssymb,amsfonts,graphicx}
\usepackage{todonotes}
\usepackage{color}
\usepackage{hyperref}
\usepackage{graphicx}

\usepackage{todonotes}   % notes showed

\shortauthors{G\"unthardt et al.}

\begin{document}

\title{The Nuclear Source of the Galactic Wind in NGC 253}

\author{G. I. G\"unthardt}
\affiliation{Observatorio Astron\'omico de C\'ordoba, Laprida 854, X5000BGR, C\'ordoba, Argentina.}

\author{D\'{i}az R.J.}
\affiliation{Observatorio Astron\'omico de C\'ordoba, Laprida 854, X5000BGR, C\'ordoba, Argentina.}

\affiliation{Gemini Observatory, 950 N Cherry Ave., Tucson, AZ85719, USA.}

\author{M. P. Ag\"uero}
\affiliation{Observatorio Astron\'omico de C\'ordoba, Laprida 854, X5000BGR, C\'ordoba, Argentina.}
\affiliation{Consejo de Investigaciones Cient\'{i}ficas y T\'ecnicas de la Rep\'ublica Argentina, Avda. Rivadavia 1917, C1033AAJ, CABA, Argentina.}

\author{G. Gimeno}
\affiliation{Gemini Observatory, 950 N Cherry Ave., Tucson, AZ85719, USA.}

\author{H. Dottori}
\affiliation{Instituto de F\'isica, Universidade Federal do Rio Grande do Sul, Brasil}

\author{J. A. Camperi}
\affiliation{Observatorio Astron\'omico de C\'ordoba, Laprida 854, X5000BGR, C\'ordoba, Argentina.}

\begin{abstract}

 We present Br$\gamma$ emission line kinematics of the nuclear region of NGC\,253, recently known to host a strong galactic wind that limits the global star formation of the galaxy. We obtained high-resolution long-slit spectroscopic data with PHOENIX at Gemini-South, positioning the slit on the nucleus Infrared Core (IRC), close to the nuclear disk major axis. The spatial resolution was 0.35$\arcsec$\,($\sim$6\,pc) and the slit length was 14$\arcsec$\,($\sim$240\,pc). The spectral resolution was $\sim$74000, unprecedentedly high for galactic nuclei observations at $\sim$2.1\,$\mu$m. 
The line profiles appear highly complex, with blue asymmetry up to 3.5$\arcsec$ away of the IRC, and red asymmetries further away to NE. Several Gaussian components are necessary to fit the profile, nevertheless a narrow and a wide ones predominate. The IRC presents kinematic widths above 700\,km\,s$^{-1}$ (FWZI), and broad component FWHM$\sim$400\,km\,s$^{-1}$, the highest detected in a nearby galaxy. At the IRC, the blue-shifted broad component displays a 90\,km\,s$^{-1}$ bump in radial velocity distribution, a feature we previously detected in molecular gas kinematics. 
The narrow component velocity dispersion ($\sim$32\,km\,s$^{-1}$) is within the expected for normal galaxies and LIRGs.
Intermediate components (FWHM$\sim$150\,km\,s$^{-1}$, red-shifted to the NE, blue-shifted to the SW) appear at some positions, as well as weaker blue (-215\,km\,s$^{-1}$) and red line wings (+300\,km\,s$^{-1}$). 
The IRC depicts a large broad-vs-narrow line flux ratio (F(B)/F(N)$\sim$1.35), and the broad component seems only comparable with those observed at very high star-forming rate galaxies. The results indicate that the IRC would be the main source of the galactic winds originated in the central region of NGC\,253.

\end{abstract}


\section{Introduction}
NGC 253 is the nearest spiral galaxy with a nuclear starburst. As a distinctive signature, it presents, like other starburst galaxies (e.g. M 82), extended outflows viewed in a variety of spectral emissions. In particular, \citep{sharp2010}, from three-dimensional spectroscopic observations, mapped the outflow of NGC 253 by using several optical emission lines. The outflow of NGC 253 has also been observed with detail in H$\alpha$ by \citep{westmoquette2011}, where due to a spectral resolution of 80-90 km s$^{-1}$, they have been able to identify multiple components in the line profiles. They obtained direct imaging and kinematic data for the southern cone between 280 and 660 pc from the nucleus and reported velocities of 100-300 km s$^{-1}$. From their kinematic studies, the authors found evidence that successive waves of outflowing gas have been ejected by different episodes of star formation. The outflow in H$\alpha$ is similar in morphology to the hot gas emitted in X-ray in scales down 20 pc \citep{strick2000}. AKARI has revealed a halo in far-infrared emission that reaches 9 kpc. Besides, this halo traces the X-ray emission region \citep{bauer2008}, suggesting the presence of dust which comes from the inner regions. \citep{kaneda2009} estimate for the dust an outflow velocity with a range of 300-2000 km s$^{-1}$, so the dust would escape from the gravitational potential of the galaxy. Also from an X-ray study of abundance ratios, \citep{mitsuishi} conclude that in the superwind, disk and halo regions, the origin of the hot interstellar gas is the same, being the abundance patterns contaminated by SN of type II, what would imply that this gas would be provided by the nuclear starburst activity. That nuclear star formation might produce cosmic rays with energies up to 10$^{18}$ eV in the superwind region \citep{romero2018}.
\citep{bolatto} determined from ALMA images that the molecular outflow rate would imply that the wind originated by the starburst activity would limit the star-forming activity, and so, the final stellar content.

Evidence presented in \citep{gunth2015} (hereafter Paper I), shows that the K-band continuum peak (or infrared core, IRC), associated  with the radio source TH7 \citep{turner1985}, would be the genuine nucleus of NGC 253. The nucleus is off-center 60 pc southwest of the symmetry center of the galactic bulge isophotes as seen in the near-infrared (NIR) spectral range. From the H$_{2}$ 2.12$\mu$m rotation curve, a kinematic residual is interpreted as evidence of outflow in the proposed nucleus while from mid-infrared images a shell-like structure around the nucleus would be revealing the outflow in the nuclear region. \citep{davidge2016} also discusses the possibility of IRC as a candidate for the nucleus of NGC 253, pointing out that a nucleus should present a substantial stellar age spread. In fact, the author finds from a spectroscopic NIR study of IRC, deep CN and C$_{2}$ bands, which are spectroscopic signatures of an underlying intermediate-age stellar population.  Detailed NIR spectroscopic observations have shown that low-mass super-massive black holes appear off-centered in some nearby spiral galaxies (e.g. M\,83, \citet{diaz2006}; NGC 6300,  \citep{gaspar2019}) and theoretical scenarios predict that they would have an important effect on the structural evolution of the host spiral galaxy \citep{emsellem2015}. 

The nucleus of NGC 253 presents high extinction in the visual band (\textit{Av}$\sim$11 mag; \citep{ontiveros2009}), consequently for a better understanding of the kinematic is necessary to perform studies in the infrared spectral region.\\ 
Until now, no detailed near-infrared observations of the outflow of NGC 253 were available for the nuclear region, so the high spectral resolution achieved by PHOENIX in addition with the proximity of NGC 253 (D$\sim$3.9 Mpc, 1$\arcsec$ = 17 pc; \citep{kara2003}), enables us to look into the very inner nuclear structure (IRC) and neighboring zones, and carry out a detailed study of the different kinematic structures.

\section{Observations and Data Reduction} \label{obs}
The observations were performed with the high spectral resolution (R$\sim$74000) echelle spectrometer PHOENIX \citep{hinkle2003} at the 8.1 m Gemini South Telescope, during a queue mode in 2009 November and 2010 February (Programs GS-2009B-Q-39 and GS-2010B-Q-64). We obtained long-slit spectra of the nuclear region of NGC 253 in the redshifted Br$\gamma$ emission line (21662 \AA). The spectra analyzed in this paper were taken with the slit (0.25$\arcsec$ = 3 pixel wide) oriented along the line joining the IRC and the knots seen in Ks band near the radio source TH2 \citep{turner1985}. The 14$\arcsec$ slit covered $\sim240$\,pc on PA 61${}^{\circ}$ (Figure 1a). This PA is coincident with the central circumnuclear disk major axis and it is close to the galaxy global major axis (PA=51${}^{\circ}$). The spatial scale of the detector is 0.085$\arcsec$ pixel$^{-1}$ and the wavelength dispersion scale is 0.098 \AA\, pixel$^{-1}$. During the observations, the weather conditions were very good with a subarcsec seeing ranging from 0.3$\arcsec$ to 0.4$\arcsec$ at the Ks band.

 The data reduction has been performed following standard reduction techniques with IRAF. The extractions of individual spectra have been made with 3 pixel width, each one separated by 2 pixels. Two sky-subtracted individual spectra were combined matching the IRC continuum emission peak, one from each Gemini Program. The total on-source exposure time was 1800s. The instrumental line profile measured from the arc comparison spectrum was FWHM $\sim$ 4 km s$^{-1}$. For the Gaussian component fitting of the Br$\gamma$ line profile we used the task NGAUSSFIT routines from IRAF.
 
The central 500 pc of the galaxy are shown in Figure 1(a) with a median seeing of 0.5$\arcsec$. This K-band image was obtained with Flamingos-2 \citep{eikenberry2008,gomez2012} at the Gemini South Telescope during a commissioning run in 2013 June 24 with a total exposure time of 18s (for more details see Paper I).

\section{Results and Discussion}\label{results}

In Figure 1(b), we display a 2D spectrum obtained with PHOENIX in the nuclear region of NGC 253, with a spatial coverage of about 185 pc. As can be seen, the kinematics of that region is highly complex. The IRC emission stands out as the main continuum source. Also, the Br$\gamma$ emission is more extended toward NE from IRC (up to 130 pc). The emission line profile and spatial distribution are in accordance with what we observed in the MIR/NIR images and lower resolution spectra obtained with Flamingos-2 (Fig. 20 at Paper I). The spectrum covers $\sim$70 \AA, being that the Br$\gamma$ emission line is as wide as 54 \AA \,($\sim$740 km s$^{-1}$) FWZI at the IRC position (Figure 2 (\textit{top})). Toward the NE of the IRC position, the FWZI is about 35 \AA\, ($\sim$480 km s$^{-1}$) while toward the SW it is even lower, 20 \AA \,($\sim$274 km s$^{-1}$). Attending to the complex nature of the line profiles, as a first approach, we determined the radial velocity distribution corresponding to the peak of the line (Figure 2 (\textit{bottom panel})). The rotation feature is clear, reflecting the presence of a circumnuclear disk that dominates the dynamics of the nuclear region. This velocity distribution strongly agrees with that observed in H$_{2}$ (Paper I). 

We compared the radial velocity gradients along the major axis obtained in the NIR regime, with those from optical studies. From Paper I, Figure 16, we derived the gradient obtained from the H$_{2}$ velocity distribution which resulted to be of $\sim$ 13 km s$^{-1}$ arcsec$^{-1}$ and the radial velocity gradient for the Br$\gamma$ velocity distribution  of the present work (Figure 2) is of $\sim$ 13.5 km s$^{-1}$ arcsec$^{-1}$. For the optical spectral range we used the H$\alpha$ radial velocity distribution (Figure 2) from \citet{camperi2011}, and the velocity gradient is of $\sim$ 8.5 km s$^{-1}$ arcsec$^{-1}$. From the work of \citet{arnaboldi95}, also in H$\alpha$, the derived gradient is of $\sim$ 10 km s$^{-1}$ arcsec$^{-1}$. The solid body appearance of the inner optical rotation curve must be due to the strong content of dust because the luminosity of the nuclear region does not support a constant density model. This velocity behavior is expected when the extinction is high. A simple calculation shows that for an optically thick edge-on disk of radius R, the shape of the observed radial velocity curve would be \textit{V(r)} $\sim$ [\textit{V$_{c}$(R)}/\textit{R}]\textit{r},  i.e., the rotation curve would appear to grow linearly in the highly obscured radial ranges \citep{diaz2000}. The observed differences between optical and NIR gradients are consistent with the wavelength dependence of the extinction effect \citep{valotto2004}. 

In Figure 3, we display a superposition of individual spectra extracted every 2 pixels ($\sim$0.17$\arcsec$). The spectra corresponding to the NE from the IRC are shown in Figure 3(a), while those corresponding to the SW are presented in Figure 3(b). We adopted the IRC as origin for relative positions, with positive values toward the SW from the IRC and negative ones toward the NE.
On the IRC position and its surroundings, the Br$\gamma$ line profile is blueshifted, however some positions toward the NE display a red asymmetry. Specifically, from -3.7$\arcsec$ toward SW, the central Br$\gamma$ line profile presents a blue asymmetry, being more pronounced at -0.5$\arcsec$, toward the NE the line profile is red asymmetric. The great resolving power of Phoenix reveals a structured Br$\gamma$ line profile. Many of them display a structure resembling steps on the blue wing of the emission line, which are more pronounced in the region between -0.9$\arcsec$ and +0.2$\arcsec$. These stepped structures are also observed in the red wing, although they have less visibility because of the steeper profile in the red side of the emission line. This behavior is more clearly seen around the IRC. As it has been mentioned before, according to \citet{westmoquette2011}, there is evidence of successive waves of outflow gas, triggered by different episodes of star formation, that could account for those observed features. Hence, multiple expanding shells integrated along the line of sight could explain the stepped profiles near the IRC. On the other hand, regarding to the southwestern spectra, it is remarkable that the red side of the emission line is almost identical for all positions independently of the line peak shifting and the intensity of the emission. This behavior could be an effect of the high extinction in this particular zone. Closer views of these profiles are displayed in Figure 3(c). These characteristics of the Br$\gamma$ line profiles are not observed in Paper I or in H$\alpha$ line profiles (\citet{westmoquette2011}), probably due to the lower spectral resolution of these observations. 

Considering the complexity of the line profiles, we modeled them using a certain number of Gaussian components with free parameters. To illustrate, in Figure 4 we show Gaussian fittings corresponding to different positions along the slit. Toward the NE from the IRC, we identified several Gaussian components, dominated by a narrow main component (hereafter ``peak'') and a broad one (hereafter ``broad''). In some spectra, it was possible to fit two less intense nearby narrow components, to the left and to the right of the main one (hereafter ``left'' and ``right''), and two more or less broad components, far from the narrow main component (hereafter ``far left'' and ``far right''). 
The far right component is, on average, +300 km s$^{-1}$ away from the emission peak, while the far left component is 215 km s$^{-1}$ blueshifted. Figures 4(a)-(c) illustrate the Br$\gamma$ line profile fittings at different positions toward the NE. In particular, the line profile at the IRC position was fitted with five components as shown in Figure 4(d). Between -4.9$\arcsec$ and -3.7$\arcsec$, where the line profile begins to show a red asymmetry, a moderate broad component is required (``intermediate''). For the most northeastern positions, the peak component is not required, being the central line profile well represented by the left and the right components. Toward SW of IRC, three components (peak, broad and left) were used to describe the line profile up to +1.2$\arcsec$, beyond that, only two components were needed, the peak component and a second one somewhat narrower than the broad component (``intermediate''). Examples of the last cases are plotted in Figure 4(e) and 4(f).

Taking advantage of the high spectral resolution, we could trace several kinematic components along the nuclear disk major axis. In Figure 5, we display the data cube projections of the line profile fitting, which involve radial velocities, FWHM and position. It may be seen that in these projections the different Gaussian components are grouped in distinctive subsystems (specially Figure 5 (\textit{top right})), giving confidence to the fitting process. There is no reference of line profiles observed with so high resolution for a starburst nucleus, making the comparison with other line profile analysis results difficult.

In Figure 5(a) (\textit{bottom}), we plotted the heliocentric radial velocity distribution corresponding to the different detached components. It is clear from the figure that the profiles are more complex in the region associated with IRC and at regions located northeast from it. The radial velocity plot shows rotational characters in almost all components. Assuming that the peak component is representative of the underlying disk kinematics at the IRC position, the rotational pattern is consistent with a Keplerian mass of (5$\pm$2) x 10$^{7}$ \textit{M}$_{\odot}$ inside 4$\arcsec$ (80 pc). Besides, the right, left and far right components rotate almost as fast as the peak component, while the far left component has a very irregular behavior and the rotational pattern is not clear. In respect to the wider components, the broad one presents a kinematic perturbation at the IRC position, making it difficult to assert a rotational pattern presence. It is always blueshifted with respect to the peak component, as illustrated in Figure 5(a) (\textit{bottom}). This character could be interpreted as an indication of a nuclear outflow originated in a star formation process (e.g. \citet{westmoquette2007c}; \citet{arribas2014}). At both sides of the positions, which exhibit a broad component, the intermediate one is noticed, being redshifted and blueshifted at the NE and SW positions respectively. It shows signs of rotation slightly slower than that of the peak component. As illustrated in Figure 5(b), these two wider components, in addition with the peak one, are the most luminous, where they are present. The broad component displays noncircular motion at the IRC position, showing an important increment of the velocity followed by an abrupt decrease. This radial velocity bump has an amplitude of $\sim$90 km s$^{-1}$ and it was already observed in the rotation curve of the molecular hydrogen obtained in Paper I. This radial velocity perturbation blurs the possible rotational pattern of the broad component. On the one hand, if a compact massive object was located at the IRC position, it could be able to distort the circular rotational pattern around it (e.g. \citet{mast2006}). Then, the observed velocity deviation from the circular motion would be accordant with an enclosed mass of 4 x 10$^{6}$ \textit{M}$_{\odot}$ (in a Keplerian approximation). On the other hand, the ejection or injection of material from or toward the IRC could be the cause of that velocity perturbation. Then, if the dominant motion were radial within the disk instead of rotational, and considering the north side of the disk as the nearest (\citet{camperi2018}, see also Figure 7 of Paper I), the bump counter-rotation would be consistent with an outflow process. Considering the presence of a blueshifted broad component around the IRC position, the mass ejection seems to be the most likely scenario to account for the bump perturbation.

The FWHM distribution is presented in Figure 5(a) (\textit{top left}). The mean FWHM for each component is as follows: (73$\pm$27) km s$^{-1}$ for the peak component, (325$\pm$24) km s$^{-1}$ for the broad component, (151$\pm$31) km s$^{-1}$ for the intermediate component, (89$\pm$27) km s$^{-1}$ for the left component, (67$\pm$24) km s$^{-1}$ for the right component, (146$\pm$45) km s$^{-1}$ for the far left, and (116$\pm$30) km s$^{-1}$ for the far right component. 
At the IRC vicinity, the peak component presents FWHM values relatively low ($\sim$60 km s$^{-1}$). At this position and its surroundings, the spectral signal-to-noise ratio (S/N) is higher, therefore we extracted spectra in every spatial pixel. The FWHM for the broad component is larger than 400 km s$^{-1}$, an unexpectedly high value for an inactive nucleus without excessive star formation.

The optical study \citep{westmoquette2007a} of the circumnuclear region of another nearby starburst galaxy, M82, involves an H$\alpha$ line profile fitting with two components showing a mean FWHM of (60$\pm$40) km s$^{-1}$ and (210$\pm$60) km s$^{-1}$ for the peak and broad components, respectively. While the FWHM of the peak component in M82 is consistent with that of our study, the FWHM of NGC 253 nuclear broad component is considerably higher than that of M82 and the highest so far detected in a nearby galaxy.

Attending to galaxies with higher star formation activity, \citet{arribas2014} studied a subsample of 11 low luminosity non-interacting LIRGs without AGN. For these so-called Class0 non-active objects, the authors studied global properties from the integrated spectra covering $\sim$9 kpc, obtaining a global velocity dispersion of (45$\pm$4) km s$^{-1}$ for the main peak component in optical emission lines. This value is higher than that of (24$\pm$5) km s$^{-1}$ expected for normal spirals \citep{epinat2010}. In the case of NGC 253, we obtained a value of (32$\pm$11) km s$^{-1}$ measured over all the positions along the slit, consistent with the moderate active star formation process in the scenario of \citet{arribas2014}. Their analysis involves only two components, therefore NGC 253 velocity dispersion of the peak component could be closer to that of LIRGS. In respect to the broad component, the subsample Class0 non-active in \citet{arribas2014} shows a FWHM average of (289$\pm$23) km s$^{-1}$, certainly comparable to that of NGC 253.

Considering mass ejection as the main broadening mechanism for the broad component, we would be interested in quantifying that phenomenon. The relative flux of the broad component in respect to the narrow one can be constrained. Around the IRC position, the broad component is more intense than the peak component with a relative flux F(B)/F(N)$\sim$1.35 (Figure 5b). However, outside IRC surroundings, the peak component dominates, F(B)/F(N)$\sim$0.72, without differences between NE and SW sides. The last value of the relative fluxes is consistent with that observed in H$\alpha$ by \citet{wood2015} in the circumnuclear region of NGC 7552, a LIRG galaxy of low intensity, and also with that for LIRGs without interacting signs (F(B)/F(N)=0.64$\pm$0.17, \citet{arribas2014}). In respect to the high values at the IRC surroundings, although \citet{wood2015} observed ratios above 1 in NGC 7552, they are not in the nuclear region but rather in regions of highest total H$\alpha$ luminosity. While some galaxies in the Class0 non-active subsample of \citet{arribas2014} show high ratios, they only observe mean relative fluxes above 1 in samples of U/LIRGs with an AGN.

 A general trend is that the velocity difference between the peak and broad components is smaller than the FWHM, so the turbulent motions would be more prominent than the bulk flows. Taking into account the spherical shell model for the outflow motion, the velocity difference between the narrow and broad components, with respect to the broad component FWHM, is the half than expected \citep{wood2015} with (V$_{broad}$ - V$_{narrow}$)/FWHM$_{broad}$ $\approx$ 0.22. In consequence, there should be another widening mechanism such as turbulent mixing layers or shocks producing the broad component width. Nevertheless, we determined the maximum velocity of the outflow as used by \citet{arribas2014} with a mean value of (237$\pm$26) km s$^{-1}$. Then, the NGC 253 nuclear outflow maximum velocity is relatively high considering that is as much as a twice of the maximum outflow velocity detected in NGC 7552 \citep{wood2015} and also higher than the mean value for the Class0 non-active subsample in \citet{arribas2014} (V$_{max}$=(166$\pm$19) km s$^{-1}$).
 
Although NGC 253 is not an infrared powerful galaxy (L$_{\textit{IR}}$=4.1 x 10$^{10}$ \textit{L}$_{\odot}$; \citet{melo2002}), a relative high star formation rate (SFR) (SFR = 1.73 $\pm$ 0.12 \textit{M}$_{\odot}$ yr$^{-1}$, \citet{bendo2015}) is taking place within  the central 20 x 10 arcsec (340 x 170 pc). The kinematic properties obtained at the IRC position and its surroundings are overall consistent with those observed in non-active LIRGs, while the high F(B)/F(N) is expected mainly in active LIRGs.  

\section{Conclusions} \label{summary}

We studied long-slit spectra around the NIR brightest source of the nuclear region of NGC 253 known as the IRC. The NIR spectroscopy allows us to penetrate deeper into the nuclear region of the nearest starburst galaxy. Additionally, the high spectral resolution enabled the study of the Br$\gamma$ emission line kinematics at the IRC and surroundings with an unprecedented detail. The complexity of the Br$\gamma$ line profiles seems unusual for a non-AGN, and we used several Gaussian components to fit the line profile observed at high spectral resolution.  However, each component is well identified in the position-FWHM-radial velocity space displayed in Figure 5, supporting a physical identity of them.
 
 In spite of the Br$\gamma$ line profile complexity, the radial velocity distribution traced by the maximum intensity of the emission line shows the same behavior as the molecular gas, with a prevalence of rotational motion. The rotation curves are well described with a linear approximation, displaying a gradient of about 13 $\pm$ 1 km s$^{-1}$ arcsec$^{-1}$. The solid body appearance of the inner rotation curve must be due to the strong content of dust in a highly inclined optically thick disk, a scenario that is supported by the comparison of the radial velocity gradients in the NIR regime, with those observed at optical spectral ranges. The observed differences between optical and NIR gradients are consistent with the wavelength dependence of the dust extinction effect.
 
 The NE region from the IRC presents several knots of star formation with high emission in Br$\gamma$. In contrast, the SW region is devoid of knots. Furthermore, the Br$\gamma$ kinematic complexity of the NE region is well marked while the SW side presents a simpler line profile simpler. Specifically, the NE line profiles present a composite shape, depicting two kinematic structures: a main bright complex portion and a pair of weak emissions well separated from the peak of the line (far left and far right components), one of them toward blue wavelength (-215 km s$^{-1}$, far left) and another toward red ones (+300 km s$^{-1}$, far right). These far components have similar profiles and present an almost constant intensity without variation with the position along the slit. On the other hand, the main portion of the line profile was described with up to four components: the narrow peak component, two moderately wide components on both sides of the peak (left and right components), and a broad or an intermediate one. The right and far right components follow the global disk rotation traced by the peak component, while the left one shows a noisy behavior probably due to the fitting of that component being highly dependent of the broad one.  On the other hand, the SW emission line profiles were fitted by only three components (peak, broad and left) at positions closer to the IRC and with only two components (peak and intermediate) further away. Although the far components are only seen NE of the IRC, we cannot exclude the possibility that, taking into account that Br$\gamma$ emission toward the SW from the IRC are less intense, these far components perhaps are not detected due to a very low S/N ratio. One possible explanation for the presence of these far components is that we are seeing an older ejection of matter. Stronger dust extinction in the southwestern side of the IRC could account for the overall morphological and kinematic differences.
 
  In addition of the broad and intermediate components being blueshifted in respect to the peak component suggesting the presence of mass ejection, a distinctive signature in the broad component radial velocity distribution is the presence of a bump perturbation observed at the IRC position (90 km s$^{-1}$ of amplitude). This feature was already observed in the H$_{2}$ rotation curve in Paper I. That molecular emission line exhibits a very symmetric profile well described with a Gaussian function (FWHM$\sim$130 km s$^{-1}$) without evidence of a broad component, whereas the Flamingos-2 Br$\gamma$ line profile shows a blueshifted broad component between -5.4$\arcsec$ and +4.7$\arcsec$. At lower spectral resolution, no other kinematical components are distinguished. Although the deviation from circular motion at the IRC position is insinuated in the Flamingos-2 Br$\gamma$ rotation curve, it is more prominent in the molecular gas one. 
  In the H$_{2}$ rotation curve, the amplitude of the velocity bump is about 30 km s$^{-1}$, considerably lower than that observed in the Phoenix Br$\gamma$ broad component. 
  This noncircular motion could be present in the Peak component velocity distribution, but it would not be greater than 20 km s$^{-1}$. Then, the broad component is more sensitive than the other Br$\gamma$ components to the physical phenomenon that causes the perturbation. 
  Supposing that the velocity bump in the central region of the radial velocity curve is originated by mass ejection from the IRC, a possible scenario is that the broad component arises in the ionized gas closer to the massive IRC.
Another peculiar characteristic of the broad component is that the FWHM$\sim$400 km s$^{-1}$ is considerably larger than that found by \citet{westmoquette2013} for the supercluster NGC 3603 (FWHM$\sim$70$\--$100 km s$^{-1}$). Assuming that this component is generated by an outflow process, the FWHM and maximum velocity are comparable with that of the global values in very intense star forming galaxies \citep{arribas2014}. In addition, the flux ratio between the broad and peak components at the IRC position are comparable with the mean ratio for galaxies with high SFR plus an AGN \citep{arribas2014}.

Several hypotheses have been proposed to explain the existence of a broad component, such as them being related to the action of stellar winds and supernovae. From optical and lower spatial resolution IFU spectroscopic studies,  \citet{westmoquette2007a,westmoquette2007b,westmoquette2009a,westmoquette2009b} infer that in the starburst galaxies NGC 1569 and M82, the broad component represents emission from highly turbulent mixing layers on the surface of denser gas clouds, set up by the impact of high-energy photons and fast-flowing winds from the massive star clusters. Since material is easily evaporated and ablated from these turbulent layers, the broad component should trace locations of mass-loading sites within the wind. This scenario could be applied to understand the IRC surroundings and account for the high FWHM observed in NGC 253. If that global phenomenon could be duplicated on a small scale, the turbulent mixing layers on the surfaces of the gas clouds around IRC could explain the large FWHM observed. In Paper I, from NIR diagnostic diagrams using H$_{2}$ emission line ratios, it was determined thatthe IRC and its surroundings exhibit the action of strong shock waves. This is consistent with the idea that shocks could play an important role in the broadening of the Br$\gamma$ emission line.

The stepped structure detected in the line profiles and the presence of several kinematic components arising at the IRC position are similar in radial velocity difference and velocity dispersion to the outflowing waves detected in the global velocity field of the ionized gas by \citet{westmoquette2011}. This points to a physical connection between the global outflowing gas and the outflowing waves detected at the IRC position. The IRC presents the highest Br$\gamma$ emission and kinematic width (FWZI above 700 km s$^{-1}$), exhibits a radial velocity bump at the broad component and twice the relative flux between the broad and peak component than in its surroundings, therefore it appears as the main present source of galactic winds in the nuclear region of NGC 253. Then, the IRC could be the source of the main gas ejections that generate the far, left, and right components observed in the Br$\gamma$ line profile along the slit.
mid-Infrared images display an arc (toward SE) and plume-like (toward SW) structures around the IRC, which could be associated with the strong winds exposed with the PHOENIX results. However, the long slit is not covering those mid-infrared structures. Consequently, a bidimensional kinematic map at high spectral resolution is mandatory for completing the outflow scenario in the inner parts of the nuclear region. \\

\begin{acknowledgments}
We gratefully acknowledge the anonymous referee for the useful suggestions that helped to improve the presentation of this study. We thank Percy Gomez for fruitful discussions. We acknowledge grant support from
CONICET (PIP 0523), ANPCyT (PICT 835), SeCyT-UNC (05/N030) and SeCyT-UNC (sigeva code 30720150100427CB). This research is based on observations obtained at the Gemini Observatory, which is operated by the Association of Universities for Research in Astronomy, Inc., under a cooperative agreement with the NSF on behalf of the Gemini partnership: the National Science Foundation (United States), National Research Council (Canada), CONICYT (Chile), Ministerio de Ciencia, Tecnolog\'{i}a e Innovaci\'{o}n Productiva (Argentina), Minist\'{e}rio da Ci\^{e}ncia, Tecnologia e Inova\c{c}\~{a}o (Brazil), and Korea Astronomy and Space Science Institute (Republic of Korea).

\end{acknowledgments}

\bibliographystyle{yahapj}
\bibliography{references}

\begin{figure*}
        \centering
        \includegraphics[width=\textwidth]{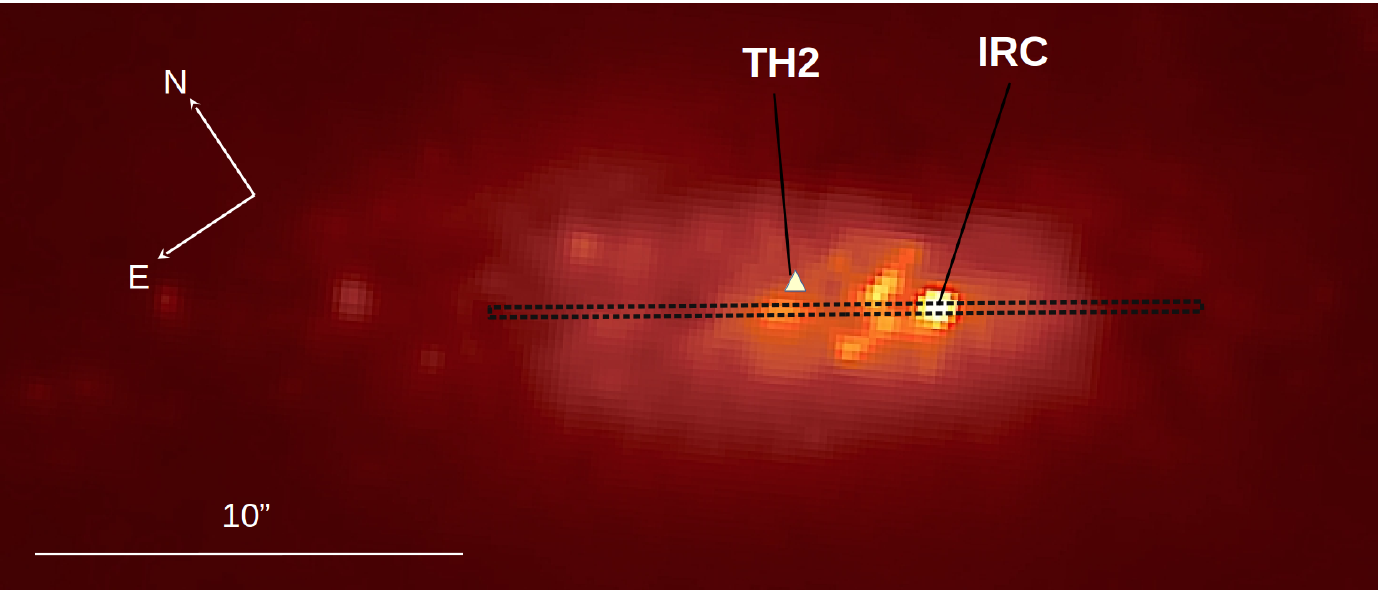}
	\includegraphics[width=\textwidth]{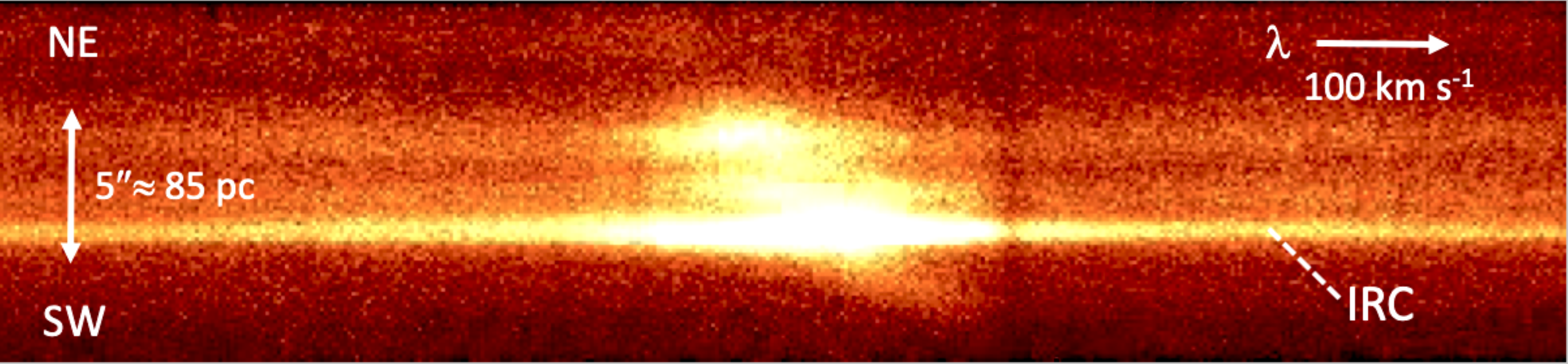}
         \caption{Top: Nuclear region of NGC 253 (Ks band) obtained with Flamingos-2. The slit is marked, and as can be seen, it covers two conspicuous regions, the most luminous knot containing IRC and the knot next to TH2. Scale: 1$\arcsec$ corresponds to 17 pc. Bottom: 2D spectrum obtained with PHOENIX at the redshifted Br$\gamma$ wavelength.}
    
\end{figure*}

\begin{figure*}
\centering 
\includegraphics[width=16cm]{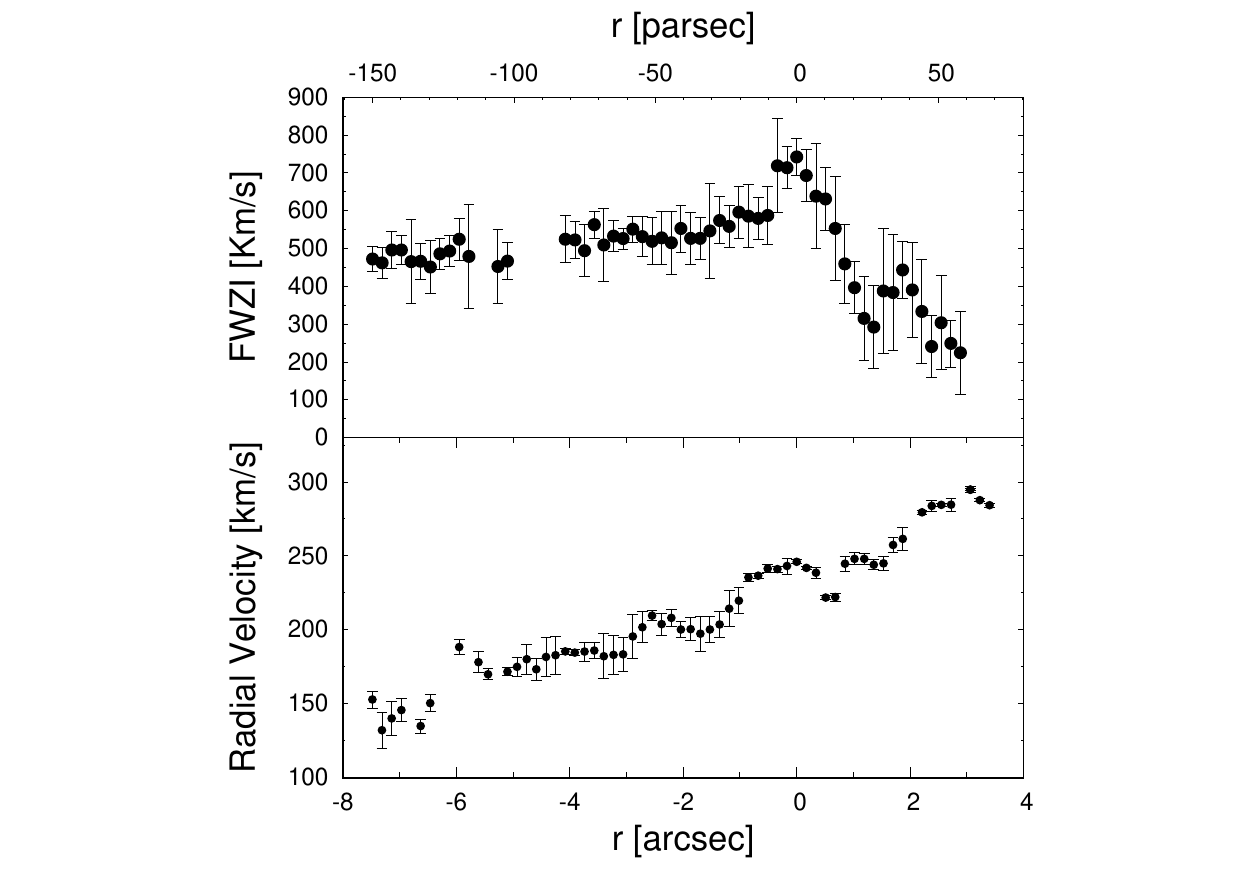} 
\caption{(\textit{Top}) Br$\gamma$ full width at zero intensity (continuum level) vs. position. The origin corresponds to the IRC position, while positive values are toward SW from it. The instrumental FWHM is $\sim$4 km s$^{-1}$. (\textit{Bottom}) radial velocity distribution (PA 61${}^{\circ}$) obtained from the maximum intensity value of the Br$\gamma$ emission line profile. }
\end{figure*}

\begin{figure*}[t]
\begin{minipage}{.45\linewidth}
\centering
\includegraphics[scale=.6]{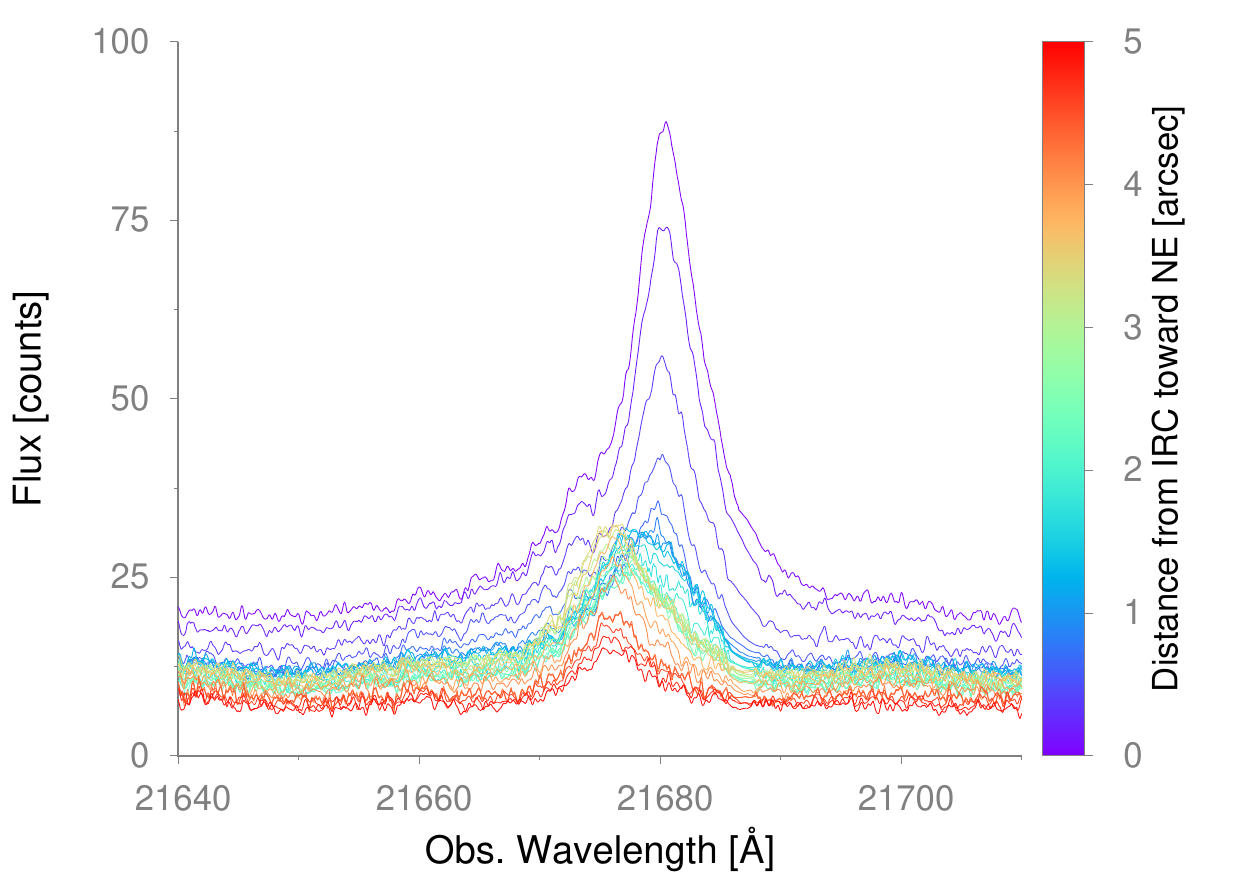}
\end{minipage}%
\begin{minipage}{.5\linewidth}
\centering
\includegraphics[scale=.6]{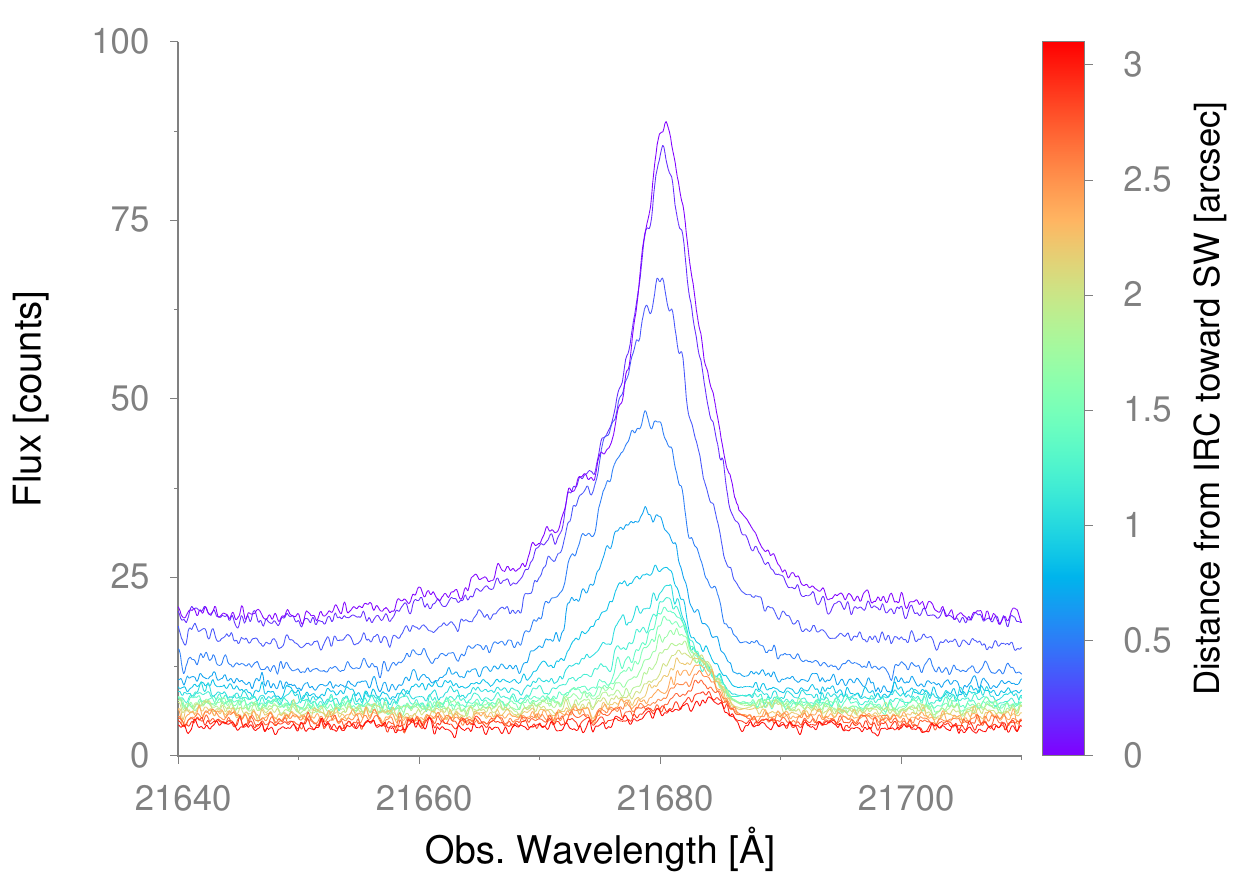}
\end{minipage}\par\medskip
\centering
\includegraphics[scale=.6]{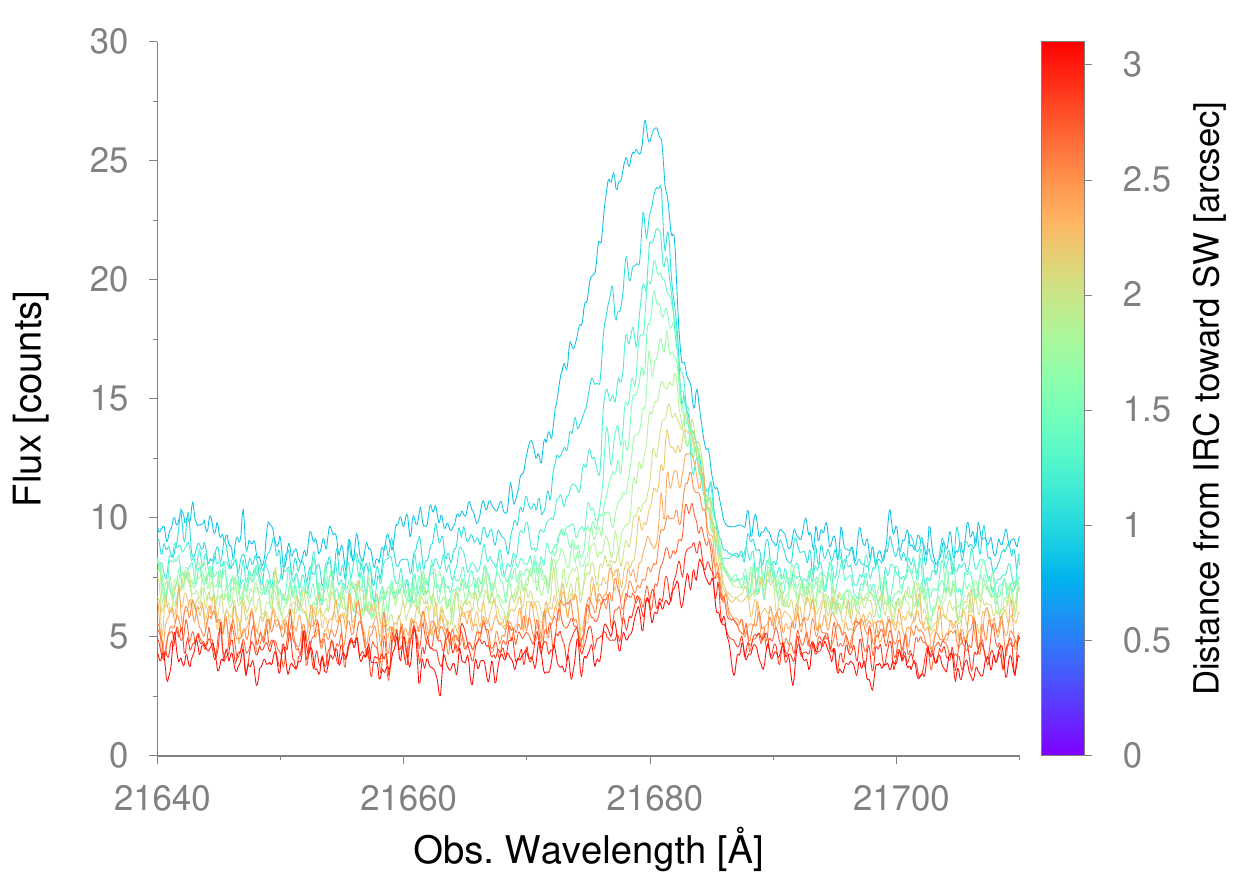}
\label{fig:main}
\caption{(a) stack of all the extracted spectra corresponding to the IRC spectrum and those northeast from it. The flux of the spectra are in arbitrary counts. (b) stack of the spectra corresponding to IRC and those southwest from it. In (c) a zoom of plot (b) is displayed, where, as mentioned in the main text, some profiles share a red wing wall between $\sim$1$\arcsec$ and 3$\arcsec$ from the IRC. In all of the plots, the color display corresponds to the distance from the IRC in arcseconds. }
\end{figure*}

\begin{figure*}
\gridline{\fig{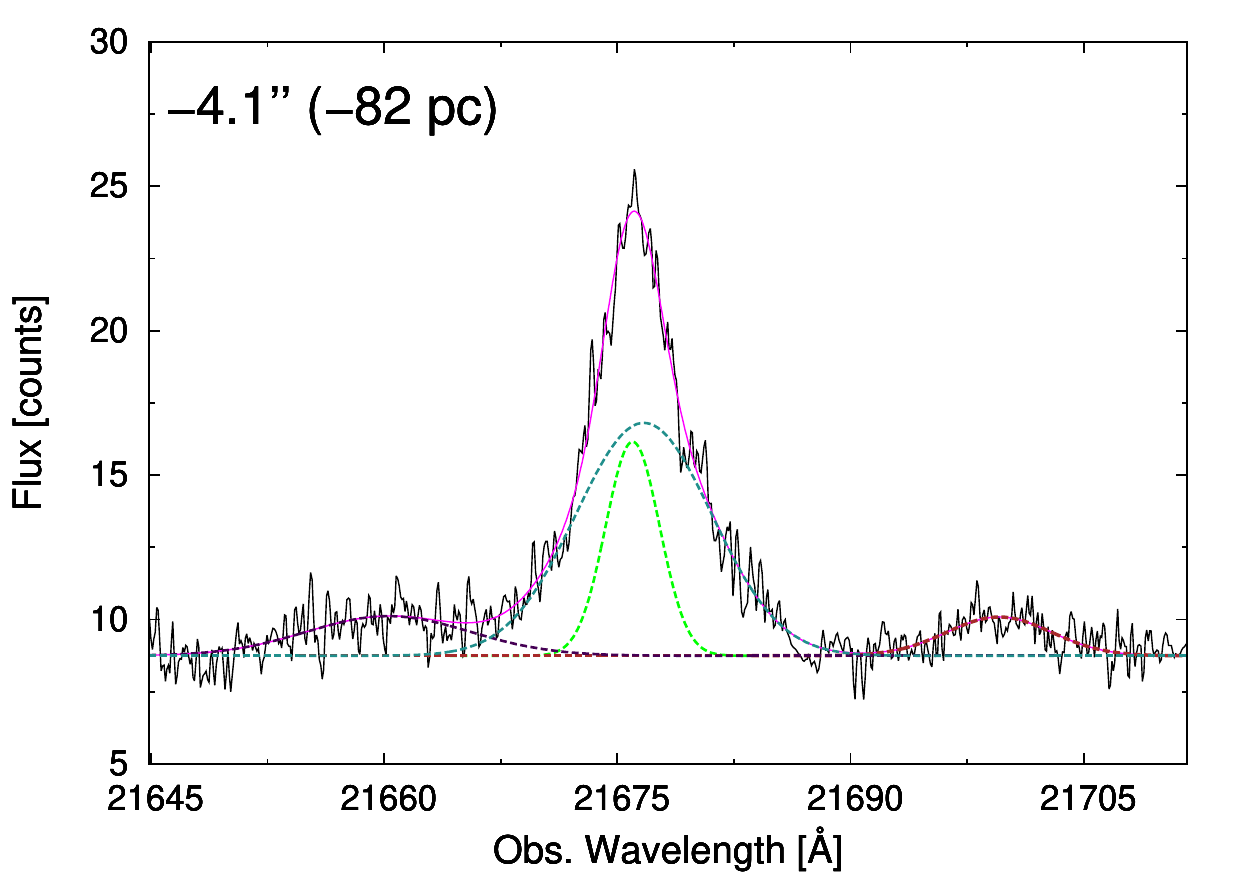}{0.4\textwidth}{(a)}
          \fig{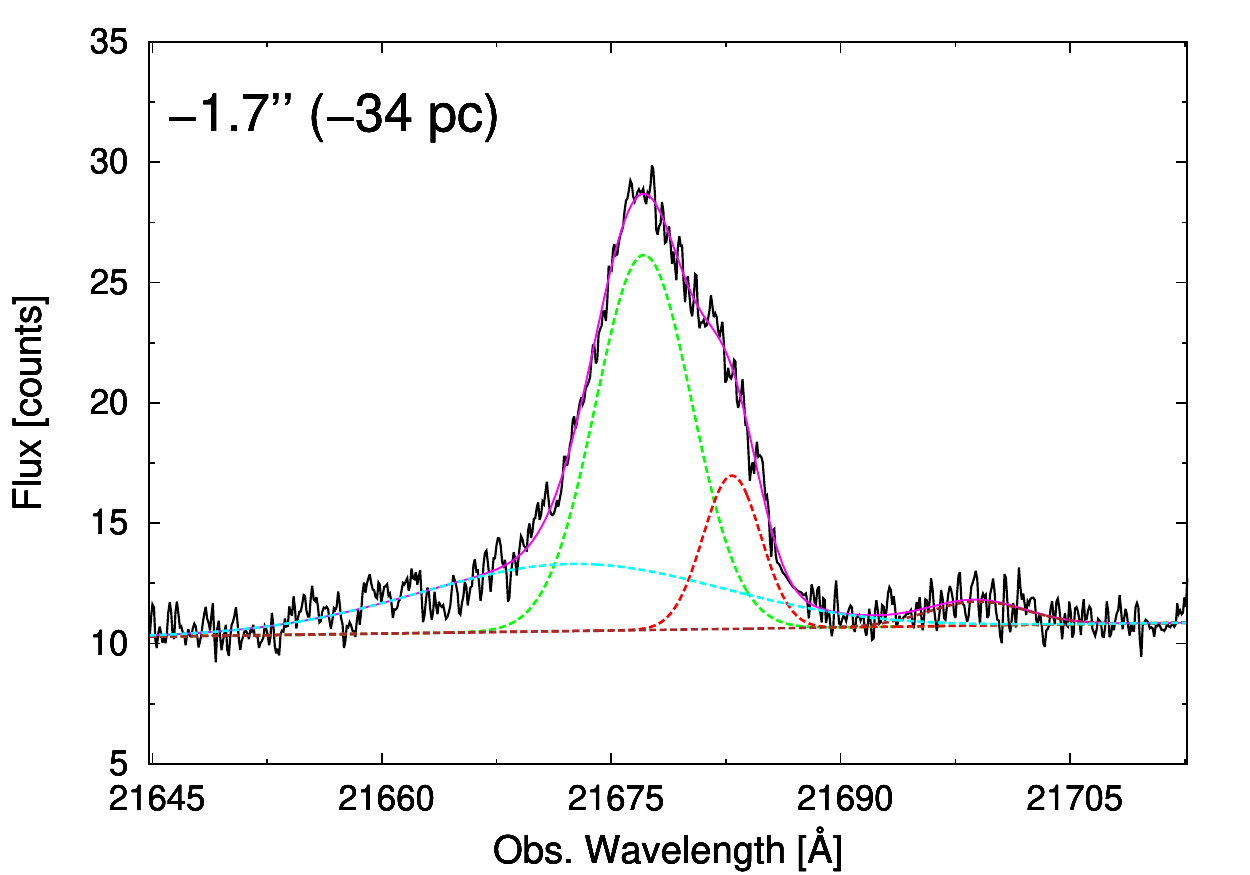}{0.4\textwidth}{(b)}}

\gridline{\fig{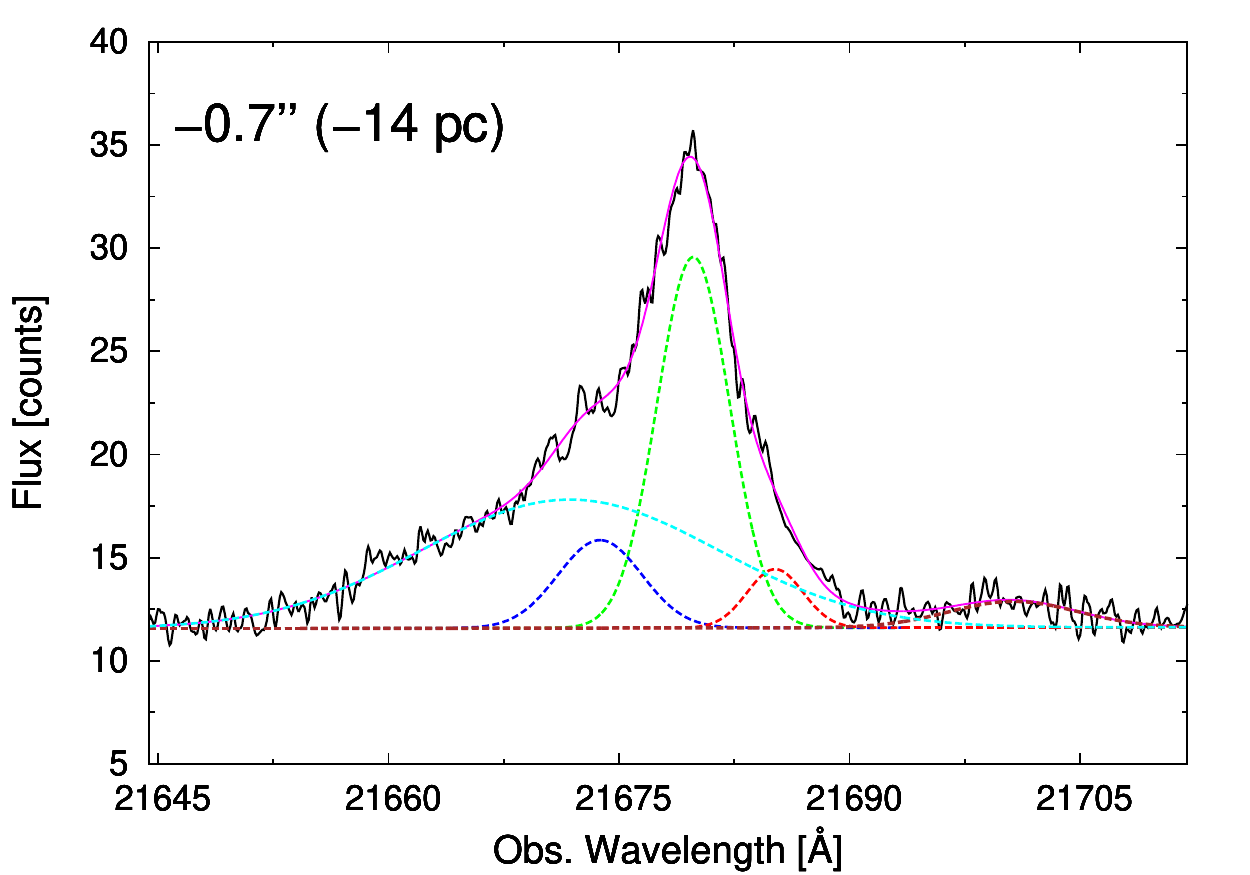}{0.4\textwidth}{(c)}          
          \fig{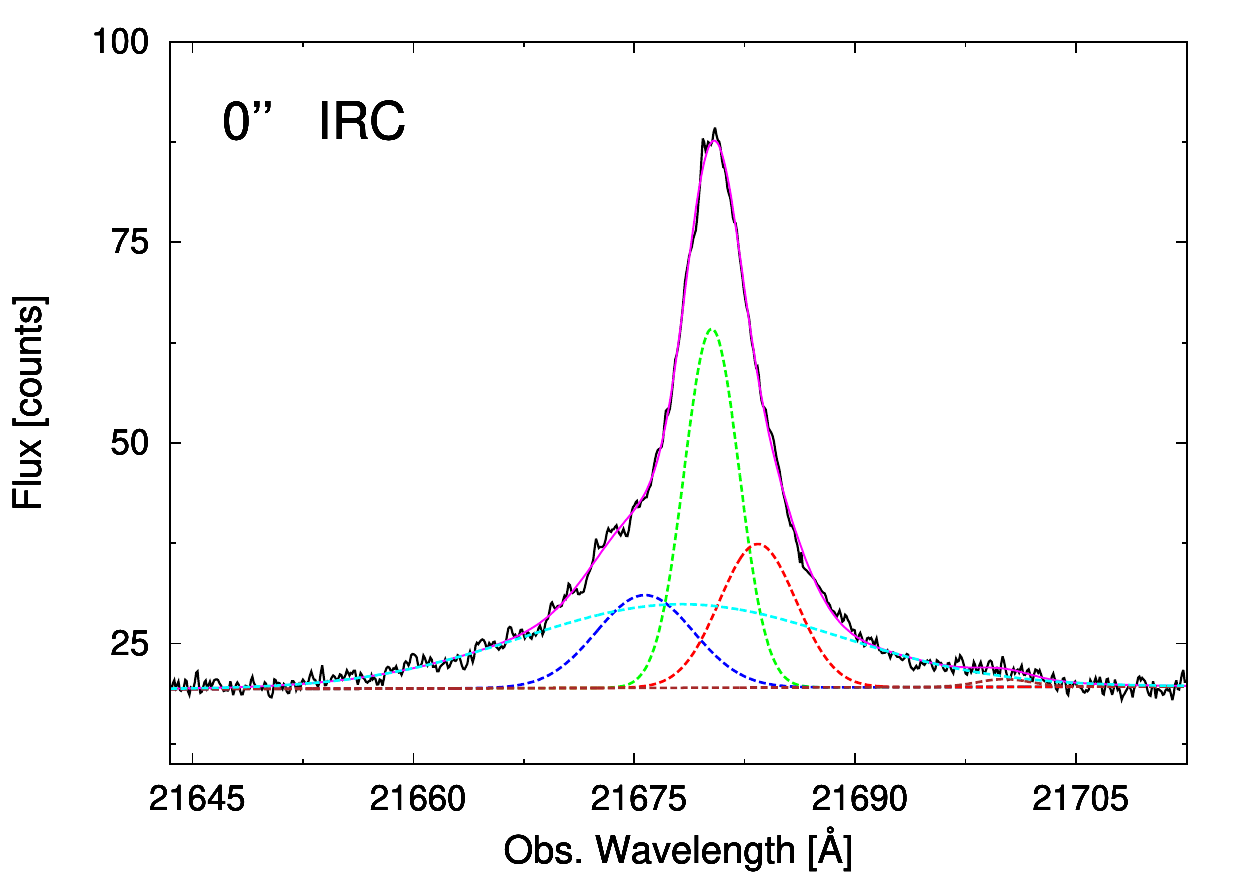}{0.4\textwidth}{(d)}}          
          
 \gridline{\fig{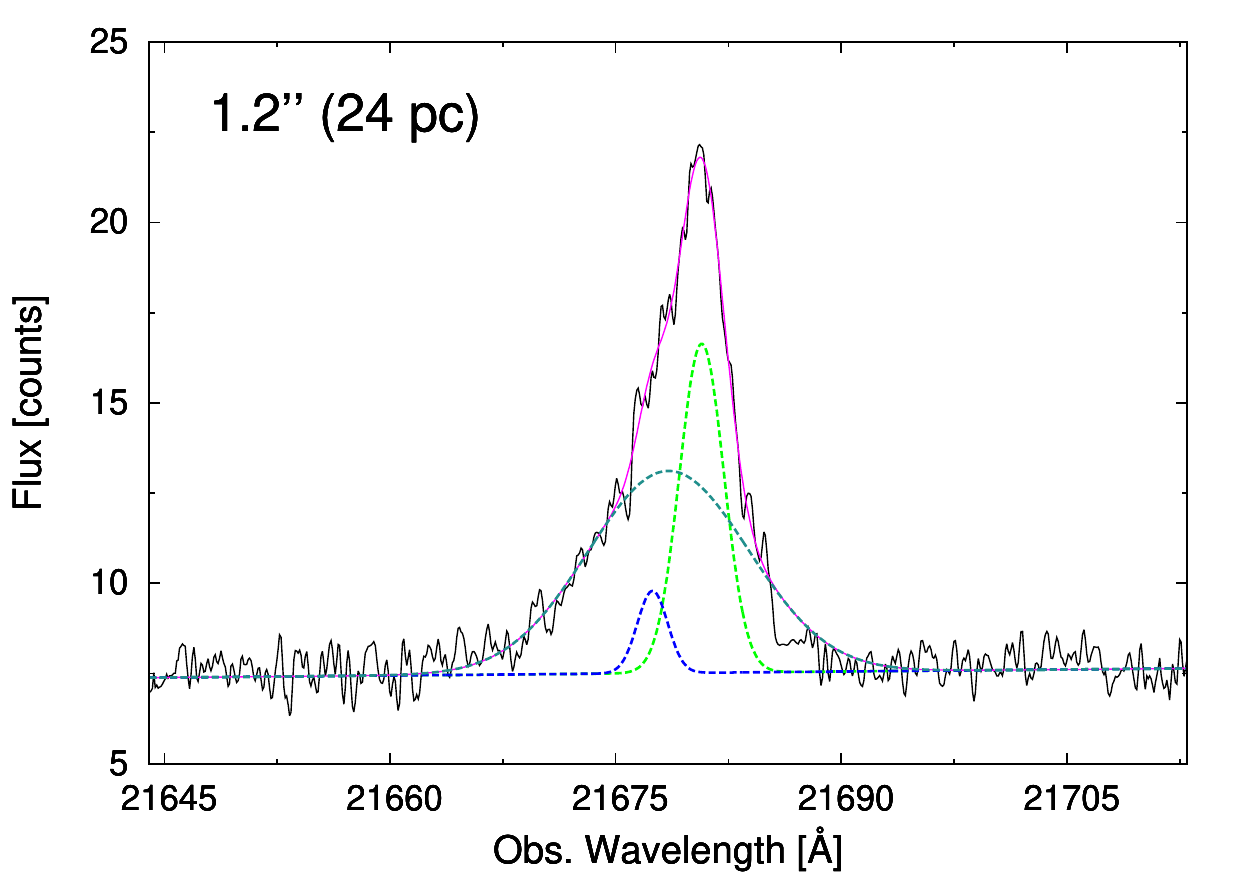}{0.4\textwidth}{(e)}
          \fig{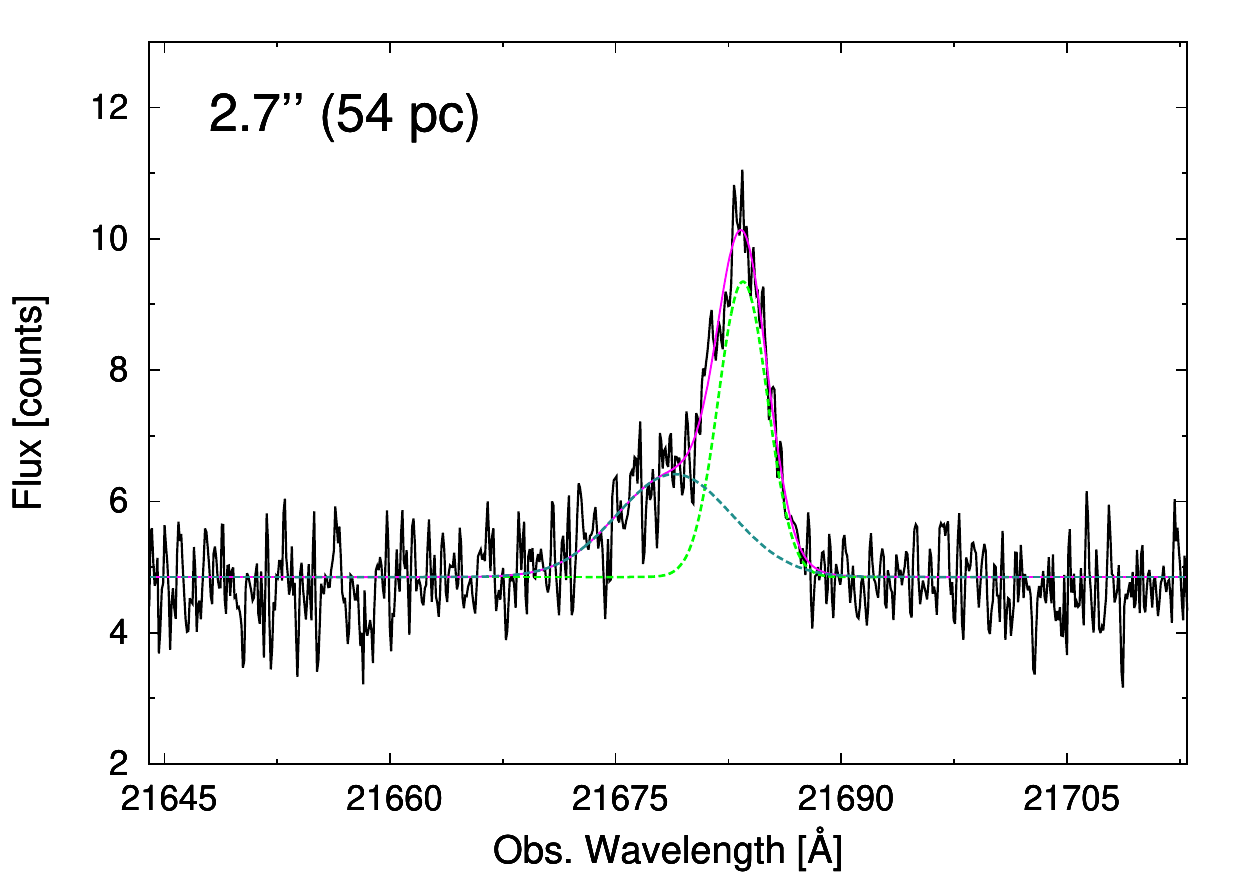}{0.4\textwidth}{(f)}}

\caption{Br$\gamma$ emission line profiles corresponding to different positions along the slit. The spectra extractions are of 0.25$\arcsec$ wide and the flux is in arbitrary counts. In each plot, at the top left corner the observed position is indicated in arcsec and between parenthesis the deprojected position in parsec. The origin corresponds to the IRC position. Negative values (three examples are shown) are associated to positions northeast from the IRC, while the two spectra southwest from the IRC are displayed at the bottom of the figure. In magenta we show the sum of the different Gaussian components. The main narrow component is in green, while the less intense neighboring components are in blue and red. The broad component is displayed in cyan, the intermediate component in dark green, while the components far from the main narrow component are plotted in purple and brown. 
}

\end{figure*}

\begin{figure*}
\centering

\includegraphics[width=18cm]{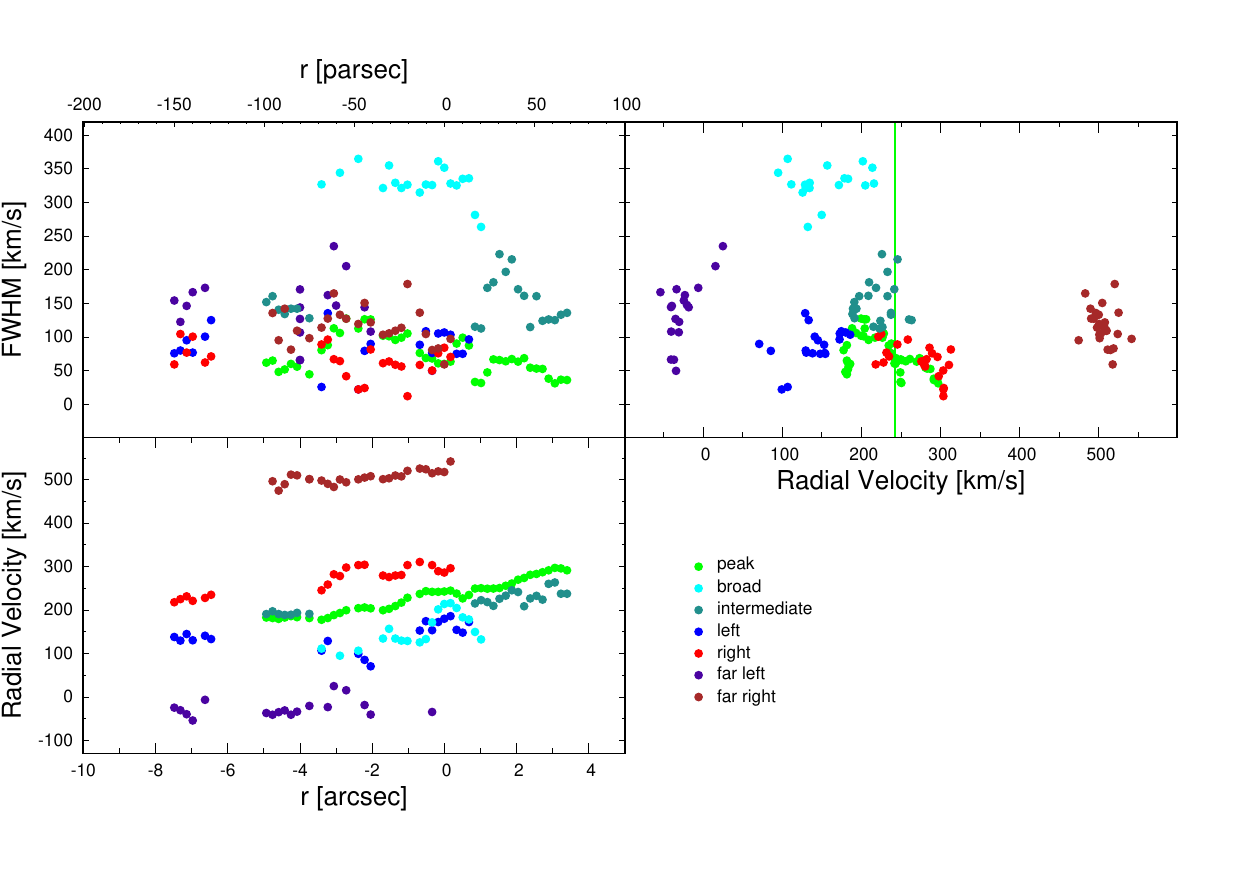} 
\includegraphics[width=10cm]{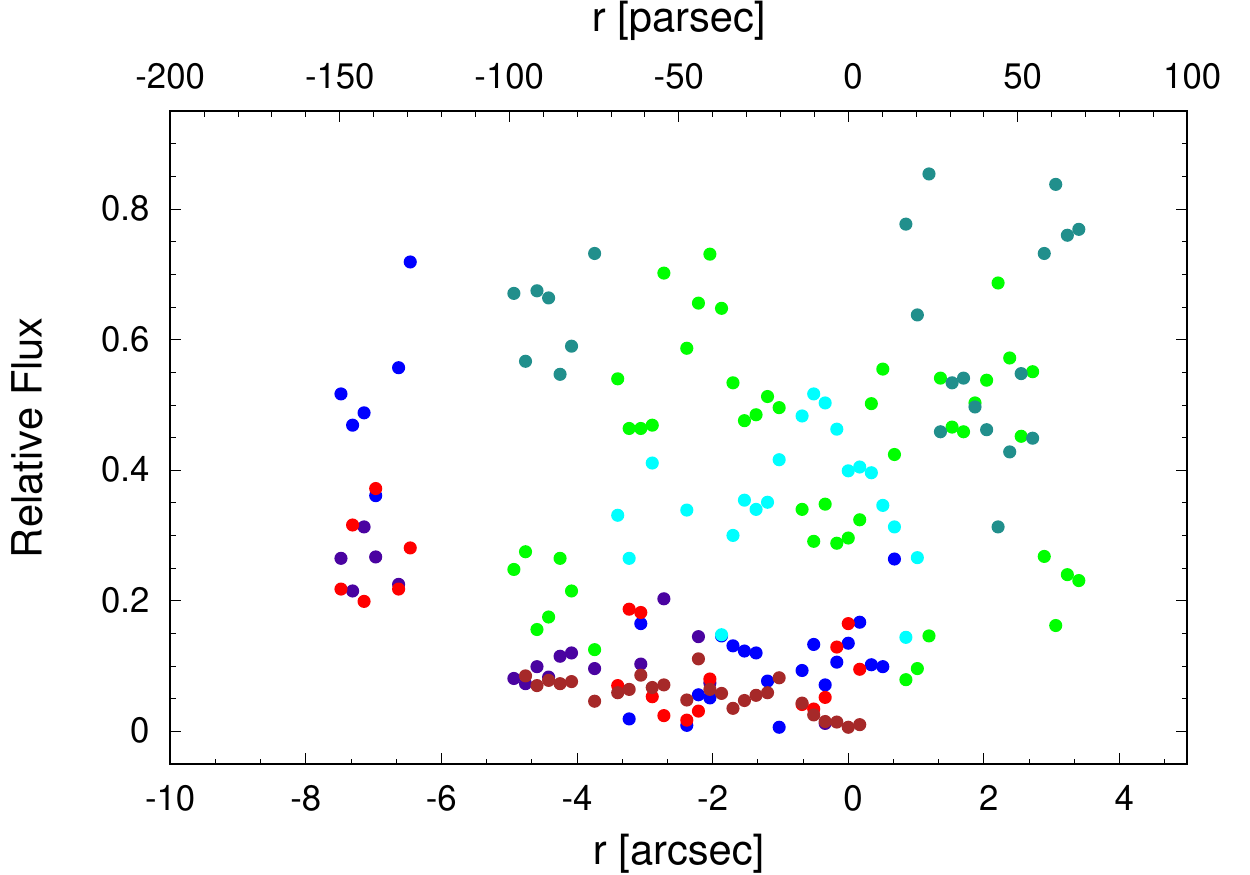} 
\caption{Top: (\textit{Top left}) FWHM distribution of the different Gaussian components fitted to the Br$\gamma$ emission line profiles. (\textit{Top right}) FWHM of the fitted components vs. radial velocity. The radial velocity corresponding to the narrow component associated to IRC is indicated as a vertical green line. (\textit{Bottom}) heliocentric radial velocity distributions associated to the different Gaussian components.  For all the plots, the narrow main component is plotted in green, their neighboring components in blue and red, the broad component is in cyan, and the intermediate component in dark green, while the farthest components are in purple and brown. Bottom: relative fluxes of the different components with respect to the total flux of the Br$\gamma$ emission line. The color symbols are the same as those in previous plots.} 
\end{figure*}

%\facility{facility ID}
%\facilities{facility ID, facility ID, facility ID} 
%\software{Numpy}

\end{document}